\documentclass[%
reprint,
 amsmath,amssymb,
 aps,
prb,
]{revtex4-1}

\usepackage{graphicx}
\usepackage{dcolumn}
\usepackage{bm}
\usepackage{cancel}

\usepackage{braket}
\usepackage[usenames,dvipsnames]{xcolor} 
\usepackage{ulem}

\begin{document}

\preprint{APS/123-QED}

\title{Theory of superconductivity in hole-doped monolayer MoS$_{2}$}

\author{Rikuto Oiwa$^{1}$}
\author{Yuki Yanagi$^{1}$}%
\author{Hiroaki Kusunose$^{1}$}%
\affiliation{%
$^{1}$Department of Physics, Meiji University, Kawasaki 214-8571, Japan
}%

\date{\today}
\begin{abstract}
\label{sec:abst}
We theoretically investigate the Cooper-pair symmetry to be realized in hole-doped monolayer MoS$_2$ by solving linearized BCS gap equations on the three-orbital attractive Hubbard-like model in the presence of the atomic spin-orbit coupling.
In hole-doped monolayer MoS$_2$, both spin-orbit coupling and the multi-orbital effects are more prominent than those of electron-doped system.
Near the valence band edge, the Fermi surfaces are composed of three different types of hole pockets, namely, one mainly consisting of the almost spin-degenerate $\ket{d_{z^{2}}}$ orbital near $\Gamma$ point, and the others of the spin-split upper and lower bands near ${\rm K}$ and ${\rm K}'$ points arising from the $\ket{d_{x^{2}-y^{2}}}$ and $\ket{d_{xy}}$ orbitals.
The number of relevant Fermi pockets increases with increase of the doping. At very low doping, the upper split bands of $\ket{d_{x^{2}-y^{2}}}$ and $\ket{d_{xy}}$ are concerned, yielding extremely low $T_{\rm c}$ due to small density of states of the split bands.
For further doping, the conventional spin-singlet state (SS) appears in the $\Gamma$ pocket, which has a mixture of the spin-triplet (orbital-singlet) (ST-OS) and spin-singlet (orbital-triplet) (SS-OT) states in the K and K$'$ pockets.
The ratio of the mixture depends on the relative strength of the interactions, and the sign of the exchange interactions.
Moderately strong ferromagnetic exchange interactions even lead to the pairing state with the dominant ST-OS state over the conventional SS one.
With these observations, we expect that the fascinating pairing with relatively high $T_{\rm c}$ emerges at high doping that involves all the three Fermi pockets.
\end{abstract}


\maketitle


\section{Introduction}
\label{sec:intro}

Noncentrosymmetric superconductors with strong spin-orbit coupling (SOC) have provided us with a new platform for investigating exotic superconductivity~\cite{Bauer,Smidman}.
The researchers have focused mainly on the Rashba-type SOC in various polar materials~\cite{Gorkov,Bauer2,Frigeri}, and it causes an in-plane helical spin texture in momentum space. More recently, Ising-type SOC have been studied as well both experimentally and theoretically~\cite{Mak, Radisavljevic, Wang, Xiao, Cao, Mak2, Zeng, Sanchez, Rose, Zeng2, Wu, Gong, Song, Majidi, Zhang, Suzuki, Wu2, Mak3, Ye, Saito_review, Lu, Saito, Barrera, Xi, Ge, Roldan, Yuan, Rosner, Honerkamp, Das, Ilic, Nakamura,Hsu,He,Sosenko} in which electron spin tends to be locked in out-of-plane direction.

The layered transition metal dichalcogenides (TMDCs) have attracted much interest as key semiconducting materials, not only for electrical and optoelectronic devices but for spin-valleytronic devices~\cite{Mak, Radisavljevic, Wang, Xiao, Cao, Mak2, Zeng, Sanchez, Rose, Zeng2, Wu, Gong, Song, Majidi, Zhang, Suzuki, Wu2, Mak3}.
In particular, MoS$_2$ has been known as a leading candidate for studying electric-field induced electron-doped superconductivity~\cite{Ye,Saito_review}, whose in-plane upper critical field $H_{\rm c2}$ is about four times larger than the Pauli limit at 1.5 K~\cite{Lu,Saito}.
The relatively high $H_{\rm c2}$ may be due to the spin-valley locking toward out-of-plane direction caused by Ising-type SOC in addition to broken spatial inversion symmetry~\cite{Saito}.

Moreover, it has been reported recently that the metallic monolayer TaS$_{2}$, whose band structure is quite similar to that of the hole-doped monolayer MoS$_{2}$, exhibits the largest in-plane $H_{\rm c2}$ among a family of the layered TMDCs~\cite{Barrera}.
As the few-layer TaS$_{2}$ having the global inversion symmetry also shows much higher $H_{\rm c2}$ than the Pauli limit, it is pointed out the importance of the spin-triplet component.
A similar tendency has also been observed in NbSe$_{2}$~\cite{Xi}, which can be viewed as heavily hole-doped monolayer MoSe$_{2}$.

So far, the superconducting state of MoS$_{2}$ has been observed only in the electron-doped system~\cite{Ye,Lu, Saito, Saito_review}, while the extensive studies to realize the hole-doped superconductivity have been hampered by either technical or some intrinsic reasons.
Nevertheless, the hole-doped superconductivity is much fascinating in contrast to the electron-doped one, since the valence bands have richer characteristics originating from Mo $d$ orbitals, and the resulting Cooper pairs are expected to have a richer variety as well~\cite{Barrera,Xi,Hsu,He,Sosenko}.
For instance, the topological superconductivity with mixed spin-singlet $d$-wave and spin-triplet $p$-wave states is discussed theoretically by the Coulomb repulsion in the slightly hole-doped MoS$_2$~\cite{Hsu}. It has also been proposed the topological superconductivity driven by the on-site attraction in NbSe$_2$ and TaS$_2$ under the magnetic field~\cite{He}.
Meanwhile, a mixture of the spin-singlet and spin-triplet states has been claimed by the on-site attraction at very low doping~\cite{Sosenko}.

So far no systematic investigations on the doping dependence have been performed, however, the doping rate is an important factor in the TMDCs, since TaS$_2$~\cite{Barrera} and NbSe$_2$~\cite{Xi} exhibit superconductivity, while MoS$_{2}$ does not, where the hole career in the formers is much larger than that in the latter.
Therefore, a systematic study of the possible pairing states on hole doping is highly desired in monolayer TMDCs. 

In this paper, we theoretically investigate the Cooper-pair symmetry in hole doping, on the basis of the three-orbital attractive Hubbard-like model.
Using the realistic tight-binding model for MoS$_{2}$ from the first-principles band calculation~\cite{Cappelluti, Liu, Zhu, Zahid, Kormanyos}, and assuming spin-independent spherical interactions as a leading mechanism for superconductivity, we obtain the (linearized) gap equations.
Solving the linearized gap equations for each irreducible representation of the $D_{3h}$ point group, we determine the Cooper-pair symmetry and $T_{\rm c}$ at various hole doping rates.

It is turned out that $T_{\rm c}$ is too low to be observed at very low doping, since the density of states (DOS) of the upper branch of the split bands around K and K$'$ points, which has the main pockets for superconductivity, is considerably small.
This may be one of the reasons why no superconductivity is observed in hole 
  doping.
However, for further dopings, the number of the relevant Fermi pockets increases showing a variety of the pairing state with higher $T_{\rm c}$.
This observation indicates that the hole-doped superconductivity could be realized at moderately high dopings.

This paper is organized as follows. 
In Sect.~II, we introduce the three-orbital attractive Hubbard-like model including the atomic SOC.
Then, we derive the (linearized) gap equations in terms of the symmetry classified gap components.
In Sect.~III, we exhibit the solutions of the linearized gap equations for several sets of interaction parameters and doping rates.
We mainly discuss the doping dependences.
The final section summarizes the paper.

\section{Model and gap equations}
\label{sec:model}

\begin{figure}[t!]
\begin{center}
\includegraphics[width=8.5cm]{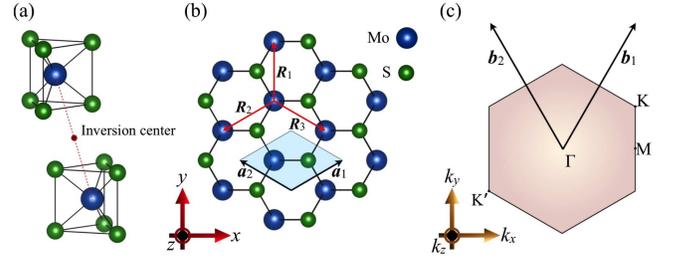}
\caption{
Crystalline structure of MoS$_2$ drawn by VESTA~\cite{vesta} and the first Brillouin zone.
(a) The unit cell of the bulk MoS$_{2}$ with the trigonal prismatic coordination. The red point represents an inversion center.
(b) The top view of monolayer MoS$_2$. The blue and green spheres represent Mo and S atoms, respectively. $\bm{R}_{1\textrm{-}3}$ and $-\bm{R}_{1\textrm{-}3}$ stand for the Mo-Mo nearest neighboring vectors, and $\bm{a}_{1}$ and $\bm{a}_{2}$ are the unit vectors as the unit length $a\equiv 1$. The diamond-shaped area indicates the two-dimensional unit cell. (c), The two-dimensional Brillouin zone. $\bm{b}_{1}$ and $\bm{b}_{2}$ are the reciprocal unit vectors.
}
\label{fig:1}
\end{center}
\end{figure}

The unit cell of the bulk MoS$_2$ consists of two units. Each unit is made up of one Mo atom located at the center of six S atoms at corners of the triangular prism, which constitutes a building block of a MoS$_2$ monolayer~\cite{Cappelluti}. The bulk MoS$_2$ has $D_{6h}^{4}$ symmetry, and there is an inversion center between two
monolayers as shown in Fig.~\ref{fig:1}(a). On the other hand, monolayer MoS$_2$ has $D_{3h}^{1}$ symmetry with lack of spatial inversion symmetry. The top view of the monolayer MoS$_{2}$ is shown in Fig.~\ref{fig:1}(b), and the corresponding Brillouin zone is shown in Fig.~\ref{fig:1}(c).

It is well known that the Bloch states of monolayer MoS$_{2}$ near the band edges consist mostly of Mo $d$ orbitals, $d_{z^{2}}$, $d_{x^{2}-y^{2}}$, and $d_{xy}$, where the contributions from $d_{yz}$, $d_{zx}$, and S $p$ orbitals are negligible~\cite{Cappelluti, Liu, Zhu}.

In this section, we first introduce the three-orbital tight-binding model with the Ising-type SOC.
Then, the Hubbard-like effective interactions are introduced under the assumption of spherical symmetry, which may arise predominantly from electron-phonon interactions among orbitals.
After setting up the model Hamiltonian, we derive the linearized BCS gap equations to be solved.

\subsection{Three-orbital tight-binding model}
\label{subsec:3tbm}

The Mo five $d$ orbitals in the trigonal prismatic structure of S atoms split into A$'_1$ ($d_{z^{2}}$), E$'$ ($d_{x^{2}-y^{2}}$, $d_{xy}$), and E$''$ ($d_{zx}$, $d_{yz}$) orbitals, where the latter two orbitals in E$''$ are far from the band edges, and hence they are omitted. In this paper, we use the magnetic quantum-number representation instead of the real one for the relevant $d$ orbitals as
\begin{align}
\Ket{0} \equiv \Ket{d_{z^{2}}},
\quad
\Ket{\pm 2} \equiv \frac{1}{\sqrt{2}}\left(\Ket{d_{x^{2}-y^{2}}}\pm i \Ket{d_{xy}}\right).
\label{eq:basis}
\end{align}

We consider the Mo-Mo nearest-neighbor hoppings, $t_{\rm A}$ for $\ket{0}$-$\ket{0}$, $t_{\rm E}$ for $\ket{\pm2}$-$\ket{\pm2}$, $t_{\rm E}'$ for $\ket{\pm2}$-$\ket{\mp2}$, and $t$ for $\ket{0}$-$\ket{\pm2}$ orbitals.
Then, the three-orbital tight-binding model measured from the chemical potential $\mu$ is given by
\begin{align}
&
\mathcal{H}_{\rm kin} = \sum_{\bm{k}\sigma}\sum_{mm'}\left(\varepsilon_{mm'}(\bm{k})-\mu\,\delta_{mm'}\right) c^{\dagger}_{\bm{k}m\sigma}c^{}_{\bm{k}m'\sigma},
\\&
\varepsilon(\bm{k}) =
\begin{pmatrix}
\varepsilon_{\rm A}+t_{\rm A}(\bm{k})  &  t^{+}(\bm{k})                           &   t^{-}(\bm{k})  \\
t^{+*}(\bm{k})                 &  \varepsilon_{\rm E}+t_{\rm E}^{+}(\bm{k})  & t_{\rm E}'(\bm{k})   \\
t^{-*}(\bm{k})                 &  t_{\rm E}'^{*}(\bm{k})                        & \varepsilon_{\rm E}+t_{\rm E}^{-}(\bm{k}) 
\end{pmatrix},
\end{align}
with the orbital basis $(\ket{0},\ket{+2},\ket{-2})$ and the spin $\sigma=\pm1$ (up/down). 
Here, $\varepsilon_{\rm A}$ and $\varepsilon_{\rm E}$ represent the energy levels of $\ket{0}$ and $\ket{\pm 2}$ orbitals, respectively, and the hopping matrix elements are given by
\begin{align}
&t_{\rm A}({\bm{k}}) = {\rm Re}\left(2t_{\rm A}\gamma_{0, \bm{k}}\right), 
\\
&t_{\rm E}^{+}(\bm{k}) = {\rm {Re}}\left(2t_{\rm E}\gamma_{0, \bm{k}}\right),
\\
&t_{\rm E}^{-}(\bm{k}) = {\rm {Re}}\left(2t^{*}_{\rm E}\gamma_{0, \bm{k}}\right),
\\
&t^{+}(\bm{k}) = t\gamma_{-1, \bm{k}}+t^{*}\gamma^{*}_{+1, \bm{k}},
\\
&t^{-}(\bm{k}) = t^{*}\gamma_{+1, \bm{k}}+t\gamma^{*}_{-1, \bm{k}},
\\
&t_{\rm E}'(\bm{k}) = t_{\rm E}'\left(\gamma_{-1, \bm{k}}+\gamma^{*}_{+1, \bm{k}}\right),
\end{align}
where we have introduced
\begin{multline}
\gamma_{n,\bm{k}} = e^{ik_{y}}+2e^{-ik_{y}/2}\cos\left(\frac{\sqrt{3}k_{x}}{2}-\frac{2\pi}{3}n\right),
\\
(n=0,\pm1).
\end{multline}
Note that according to symmetry arguments, it is shown that the hopping integrals $t_{\rm A}$ and $t_{\rm E}'$ are real, while $t_{\rm E}$ and $t$ are complex.
The imaginary parts of $t_{\rm E}$ and $t$ arise from the hoppings between $\ket{d_{x^{2}-y^{2}}}$ and $\ket{d_{xy}}$ orbitals, and between $\ket{d_{xy}}$ and $\ket{d_{z^{2}}}$ orbitals, respectively, by implicitly taking account of the presence of S atoms. 
Otherwise the imaginary parts of $t_{\rm E}$ and $t$ vanish and the system becomes the triangular lattice having the spatial inversion symmetry.
Throughout this paper, we set $\varepsilon_{\rm A}=1.046$, $\varepsilon_{\rm E}=2.104$, $t_{\rm A}=-0.184$, $t_{\rm E}'=-0.0805$, $t_{\rm E}=0.138-0.338i$, $t=0.359+0.284i$ eV, which are taken from Ref.~\onlinecite{Liu} where the real representation for $d$ orbitals, i.e., the r.h.s. of Eq.~(\ref{eq:basis}), are used.

The atomic SOC is given by
\begin{equation}
\mathcal{H}_{\rm SOC} = \frac{\lambda}{2}\sum_{\bm{k}m\sigma}(m\sigma) c^{\dagger}_{\bm{k}m\sigma}c^{}_{\bm{k}m\sigma},
\end{equation}
where $\lambda$ is the strength of the SOC, which is estimated from the direct gap of the band calculation as $\lambda \sim 0.073$ eV~\cite{Liu}, and we adopt this value. 
$\mathcal{H}_{\rm SOC}$ has only the diagonal matrix elements, i.e., Ising-type, since we consider the orbitals with $m=0$, $\pm2$ which are connected only by quadrupolar transitions.
As a result, the $z$-component of the spin $\sigma$ becomes a good quantum number.

\begin{figure}[t]
\begin{center}
\includegraphics[width= 1.0 \hsize]{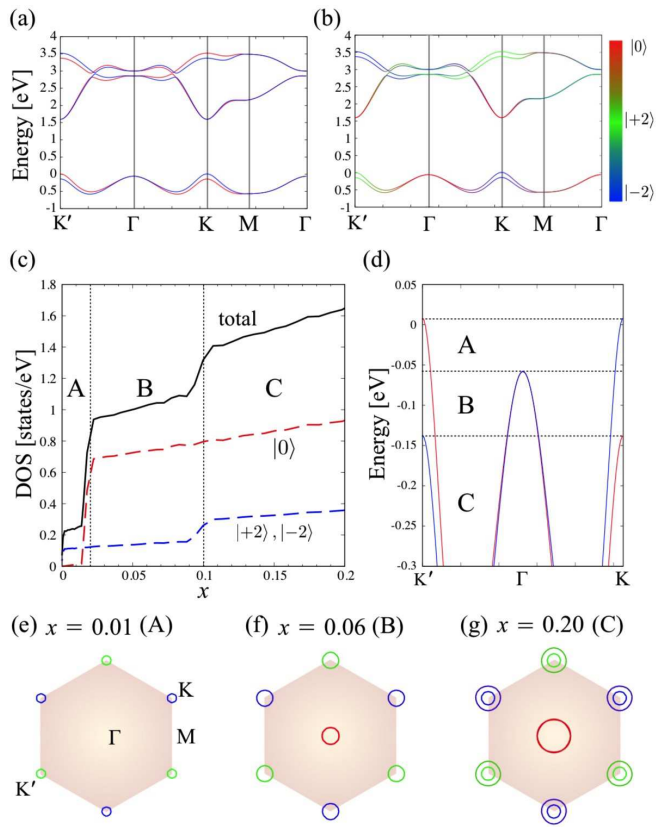}
\caption{
Band structure, the DOS, and the Fermi surfaces of monolayer MoS$_2$.
(a) The spin dependence. The red and blue lines represent up- and down-spin bands, respectively.
(b) The orbital dependence, where the red, green, and blue lines represent $\ket{0}$, $\ket{+2}$, and $\ket{-2}$, respectively.
(c) The doping dependence of the DOS at the Fermi level. 
The doubly stepwise behavior appears when the Fermi level goes across the top of the split bands as shown in (d) the enlarged view of (a).
(e)-(g) The Fermi surfaces at the representative doping rates in the A-C regions.
}
\label{fig:2}
\end{center}
\end{figure}

Using the above model parameters, the obtained energy dispersions near band edges well reproduce the essential features of the band structures from the first-principles band calculation~\cite{Liu} as shown in Fig.~\ref{fig:2}.

The spin and orbital dependences of the energy bands are shown in Figs.~\ref{fig:2}(a) and (b), respectively.
It is shown that the bottoms of the conduction bands near K and K$'$ points are almost spin degenerate, because they are mainly composed of the $\ket{0}$ orbital, and the SOC affects only through the valence bands perturbatively. 

On the other hand, the tops of the valance bands near K and K$'$ points consist predominantly of the $\ket{\pm2}$ orbitals, showing considerably large spin splitting.
On the contrary, the top of the valence bands near $\Gamma$ point is almost spin degenerate, which arises mainly from the $\ket{0}$ orbital.
Moreover, the effective mass of the latter is considerably larger than those of the former.
Reflecting the difference of the effective masses, the DOS of the $\ket{0}$ orbital is much larger than those of the $\ket{\pm2}$ orbitals.

The characteristic features of the valence bands lead to doubly stepwise behavior in the doping dependence of the DOS at the Fermi level as shown in Fig.~\ref{fig:2}(c), where the doping $x$ is defined as $x=2-n$ with $n$ being the electron density.
It is natural to expect that the coupled orbital-spin degrees of freedom of the valence bands give rise to a variety of superconducting states in the hole doping.
The doubly stepwise changes of the pairing states in the hole doping are also expected depending on the changes of the Fermi-surface topology as shown in Figs.~\ref{fig:2}(e)-(g).

\subsection{BCS interactions}

Next, let us introduce the effective interactions leading to superconductivity.
The effective interactions may arise predominantly from the electron-phonon interactions among orbitals.
The Coulomb repulsion may also be important especially in the exchange interactions.
In the present study, we assume that the local interactions are spin independent and spherically symmetric.
With these assumptions, the pairing interactions are expressed in the form,
\begin{align}
H_{\rm int}&=\frac{1}{2N_{0}}\sum_{\bm{k}\bm{k}'\sigma\sigma'}\sum_{m_{1}m_{2}m_{3}m_{4}}V_{m_{1}m_{2},m_{3}m_{4}}
\cr&\quad\quad\quad\times
c^{\dagger}_{\bm{k}m_{1}\sigma}c^{\dagger}_{-\bm{k}m_{2}\sigma'}c^{}_{-\bm{k}'m_{4}\sigma'}c^{}_{\bm{k}'m_{3}\sigma},
\end{align}
where $N_{0}$ is the number of the unit cells.
The matrix elements are finite at least for $m_{1}+m_{2}=m_{3}+m_{4}$, and they satisfy the relation, $V_{m_{1}m_{2}m_{3}m_{4}}=V_{m_{3}m_{4}m_{1}m_{2}}=V_{m_{2}m_{1}m_{4}m_{3}}=V_{-m_{1}-m_{2}-m_{3}-m_{4}}$.
Due to the spherical symmetry, they are parameterized as
\begin{align}
&\quad
V_{\pm2\mp2\pm2\mp2}=U-J_{2},
\quad
V_{\pm2\mp2\mp2\pm2}=2J_{2},
\cr&\quad
V_{0000}=U,
\quad
V_{\pm2\mp200}=J_{0},
\quad
(m_{1}+m_{2}=0),
\\&\quad
V_{\pm2\pm2\pm2\pm2}=U-J_{2},
\quad
(m_{1}+m_{2}=\pm4),
\\&\quad
V_{0\pm20\pm2}=V_{\pm20\pm20}=U-2J_{0},
\cr&\quad
V_{0\pm2\pm20}=V_{\pm200\pm2}=J_{0},
\quad
(m_{1}+m_{2}=\pm2).
\label{int2}
\end{align}
$U$ is the direct interaction, while $J_{0}$ and $J_{2}$ are the exchange interactions between the $\ket{0}$ and $\ket{\pm2}$ orbitals and among the $\ket{\pm2}$ orbitals, respectively.
In contrast to the $1/r$ Coulomb interaction, the positiveness of $U$, $J_0$, and $J_2$ are not guaranteed~\cite{Nomura}.
In this paper, we assume that $U<0$ is the leading attractive interaction for superconductivity, and $J_{0}=J_{2}\equiv J$ for simplicity.
Since the Coulomb repulsions may contribute to the exchange interactions in addition to the electron-phonon attractions, both cases, $J>0$ (ferromagnetic) and $J<0$ (antiferromagnetic), are discussed.

Finally, the BCS Hamiltonian to be solved in this paper is given by
\begin{align}
\mathcal{H}_{\rm BCS}=\mathcal{H}_{\rm kin}+\mathcal{H}_{\rm SOC}+\mathcal{H}_{\rm int}.
\end{align}

\subsection{Gap equations}
\label{subsec:gap}

In order to discuss superconducting symmetry in the hole doping, let us introduce the isotropic ($s$-wave) gap function,
\begin{align}
&\Delta_{mm'}^{\sigma\sigma'} = - \frac{1}{N_0}\sum_{\bm{k}}\sum_{m''m'''}V_{mm'm''m'''}
\Braket{c_{\bm{k}m''\sigma}c_{-\bm{k}m'''\sigma'}}.
\label{gapeq}
\end{align}

Since the Fermi surfaces are mainly composed of the $\ket{0}$ orbital near $\Gamma$ point, and the $\ket{\pm2}$ orbitals near K and K$'$ points, we consider the Cooper pairs within the $\ket{0}$ orbital, and the $\ket{\pm2}$ orbitals.
Thus, the interactions in Eq.~(\ref{int2}) are irrelevant.
The spin state of the pair in the $\ket{0}$ orbital must be spin-singlet due to anti-commutation relation, while the pairs in the $\ket{\pm2}$ orbitals are either spin-singlet (orbital-triplet) or spin-triplet (orbital-singlet).

To decompose the gap function into each components, we introduce the following matrices for the orbital sectors as
\begin{align}
&
\tau^{x}=\begin{pmatrix} 0 & 0 & 0 \\ 0 & 0 & 1 \\ 0 & 1 & 0 \end{pmatrix},
\tau^{y}=\begin{pmatrix} 0 & 0 & 0 \\ 0 & 0 &-i \\ 0 & i & 0 \end{pmatrix},
\tau^{z}=\begin{pmatrix} 0 & 0 & 0 \\ 0 & 1 & 0 \\ 0 & 0 &-1 \end{pmatrix},
\cr&
\tau^{0}=\begin{pmatrix} \sqrt{2} & 0 & 0 \\ 0 & 0 & 0 \\ 0 & 0 & 0 \end{pmatrix},
\tau^{2}=\begin{pmatrix} 0 & 0 & 0 \\ 0 & 1 & 0 \\ 0 & 0 & 1 \end{pmatrix}.
\end{align}
Then, the spin-singlet (SS) pair in the $\ket{0}$ orbital is given by
\begin{align}
\psi\frac{1}{\sqrt{2}}\tau^{0}(i\sigma_{y}),\quad\text{(SS)}.
\end{align}
Similarly, the spin-singlet (orbital-triplet) (SS-OT) and spin-triplet (orbital-singlet) (ST-OS) pairs in the $\ket{\pm2}$ orbitals are given by
\begin{align}
&
\bm{D}\cdot(i\bm{\tau}\tau^{y})(i\sigma_{y}),\quad\text{(SS-OT)},
\\&
\bm{d}\cdot(i\tau^{y})(i\bm{\sigma}\sigma_{y}),\quad\text{(ST-OS)}.
\end{align}

Introducing the 7-component d-vector,
\begin{align}
\bm{\mathcal{D}}=[\psi/\sqrt{2},\,D_{z},\,d_{z},\,d_{x},\,d_{y},\,D_{x},\,D_{y}],
\end{align}
and the corresponding 7-component expansion basis,
\begin{align}
\bm{X}=&\bigl[\tau^{0}(i\sigma_{y}),\,(i\tau^{z}\tau^{y})(i\sigma_{y}),\,(i\tau^{y})(i\sigma_{z}\sigma_{y}),\,
\cr&\quad\quad
(i\tau^{y})(i\sigma_{x}\sigma_{y}),\,(i\tau^{y})(i\sigma_{y}\sigma_{y}),\,
\cr&\quad\quad
(i\tau^{x}\tau^{y})(i\sigma_{y}),\,(i\tau^{y}\tau^{y})(i\sigma_{y})\bigr], \label{eq_x}
\end{align}
we express compactly the decomposition as
\begin{align}
\Delta_{mm'}^{\sigma\sigma'}=\bm{\mathcal{D}}\cdot\bm{X}_{mm'}^{\sigma\sigma'}.
\label{gap_decomp}
\end{align}
Using the orthonormal relation,
\begin{align}
{\rm Tr}(X^{i}X^{j\dagger})=4\delta_{ij},
\end{align}
we obtain $\mathcal{D}^{i}={\rm Tr}(\Delta X^{i\dagger})/4$. All the Cooper pair components and their symmetry in the $D_{3h}$ point group are summarized in Table~\ref{tab:1}.

\begin{table}[t!]
\begin{center}
\caption{
Symmetry of the Cooper pairs in the $D_{3h}$ point group.
``irrep.'' (``o-irrep.'') represents the irreducible representation for the spin-orbital (orbital) space.
In the presence of the Ising-type SOC, the components are classified by ``irrep.'', while in the absence of the Ising-type SOC, the symmetry operations are defined separately in the orbital and spin spaces, and the components are classified by ``o-irrep.'' and the spin magnitude $S$.
}
\label{tab:1}
\begin{tabular}{rcccccc}
\hline\hline
$i$\,\,\, & type & irrep.   & o-irrep. & $S$  & component       & basis $X^{i}$ \\ \hline
1\,\,     & SS   & ${\rm A}'_{1}$ & ${\rm A}'_{1}$       & $0$ & $\psi$         & $i\tau^{0}\sigma_{y}$  \\
2\,\,     & SS-OT &              &            &  & $D_{z}$         & $i\tau^{x}\sigma_{y}$ \\
3\,\,     & ST-OS &   & ${\rm A}'_{2}$     & $1$ & $d_{z}$         & $i\tau^{y}\sigma_{x}$ \\ \hline
4,5       & ST-OS & $\mathrm{E}''$ & ${\rm A}'_{2}$  & $1$  & $(d_{x},d_{y})$ & $(-i\tau^{y}\sigma_{z},-\tau^{y}\sigma_{0})$ \\
\hline
6,7       & SS-OT & $\mathrm{E}'$ & $\mathrm{ E}'$ & $0$  & $(D_{x},D_{y})$ & $(-i\tau^{z}\sigma_{y},-\tau^{2}\sigma_{y})$ \\ \hline\hline
\end{tabular}
\end{center}
\end{table}

A similar decomposition is also made for the pairing interaction as
\begin{align}
V_{m_{1}m_{2},m_{3}m_{4}}\delta_{\sigma_{1}\sigma_{3}}\delta_{\sigma_{2}\sigma_{4}}
=\sum_{ij}v_{ij}(X^{i})^{\sigma_{1}\sigma_{2}}_{m_{1}m_{2}}(X^{j\dagger})^{\sigma_{4}\sigma_{3}}_{m_{4}m_{3}}.
\label{int_decomp}
\end{align}
and the finite components are given by
\begin{align}
&
v_{11}=\frac{U}{4},
\quad
v_{22}=\frac{1}{4}(U+J_{2}),
\quad
v_{12}=v_{21}=\frac{J_{0}}{2\sqrt{2}},
\label{v12}
\\&
v_{33}=v_{44}=v_{55}=\frac{1}{4}(U-3J_{2}),
\label{v345}
\\&
v_{66}=v_{77}=\frac{1}{4}(U-J_{2}),
\end{align}
which tell us the strength of the interaction for each pairing state.
It is clearly shown that when the exchange interaction is ferromagnetic $J>0$ (antiferromagnetic $J<0$), the $d_z$ paring is enhanced (suppressed), while the $D_z$ paring is suppressed (enhanced).

With these preliminaries, let us discuss the linearized gap equations.
To this end, we introduce the Matsubara Green's function matrix as
\begin{align}
[G(\bm{k},i\omega_{n})]^{\sigma\sigma'}_{mm'}=-\int_{0}^{\beta}d\tau\,e^{i\omega_{n}\tau}\braket{T_{\tau}c^{}_{\bm{k}m\sigma}(\tau)c^{\dagger}_{\bm{k}m'\sigma'}},
\end{align}
where $\omega_{n}=(2n+1)\pi/\beta$ is the fermionic Matsubara frequency, and it implicitly depends on the gap function.
Then, the pair wave function in the gap function, Eq.~(\ref{gapeq}), is expressed as
\begin{multline}
\braket{c_{\bm{k}m\sigma}c_{-\bm{k}m'\sigma'}}
=T\sum_{n}[G^{(0)}(\bm{k},i\omega_{n})\Delta
\\\times
G^{\rm T}(-\bm{k},-i\omega_{n})]_{mm'}^{\sigma\sigma'},
\label{pairwf}
\end{multline}
where the superscript $(0)$ means that $G$ is evaluated in the normal state, $\Delta=0$.
It is explicitly given by
\begin{align}
G^{(0)}(\bm{k},i\omega_{n})=[i\omega_{n}+\mu-\varepsilon(\bm{k})-\lambda \tau^{z}\sigma_{z}]^{-1}. 
\label{eq_G0}
\end{align}

Note that $G^{(0)}(\bm{k},i\omega_{n})$ is diagonal in spin indices.

By approximating $G$ with $G^{(0)}$ in the pair wave function (\ref{pairwf}), we obtain the linearized gap equation.

Using the decompositions for the gap function, Eq.~(\ref{gap_decomp}), and the interaction, Eq.~(\ref{int_decomp}), we finally obtain the linearized gap equations as
\begin{align}
\mathcal{D}_{i}=\sum_{j}\chi_{ij}^{(0)}\mathcal{D}_{j},
\quad
\chi_{ij}^{(0)}=-\sum_{k}
v_{ik}K_{kj}^{(0)},
\end{align}
with
\begin{align}
K_{ij}^{(0)}=\frac{T}{N_{0}}\sum_{\bm{k}n}{\rm Tr}\left[
X^{i\dagger}G^{(0)}(\bm{k},i\omega_{n})X^{j}G^{(0)\rm T}(-\bm{k},-i\omega_{n})
\right],
\end{align}
where T represents the transpose of $G^{\rm (0)}(\bm{k},i\omega_{n})$ with respect to indices $(m,\sigma)$.

Using the eigenstate of $\mathcal{H}_{0}=\mathcal{H}_{\rm kin}+\mathcal{H}_{\rm SOC}$ and the unitary matrix which diagonalizes $\mathcal{H}_{0}$, i.e., $\mathcal{H}_{0}\ket{\bm{k}\alpha\sigma}=\xi_{\bm{k}\alpha\sigma}\ket{\bm{k}\alpha\sigma}$, and $\ket{\bm{k}m\sigma} = \sum_{\alpha}[U_{\sigma}(\bm{k})]_{m\alpha}\ket{\bm{k}\alpha\sigma}$, the Matsubara summation can be carried out, and we obtain
\begin{align}
K_{ij}^{(0)}&=\frac{1}{N_{0}}\sum_{\bm{k}\sigma_{1}\sigma_{2}\alpha\beta}
I_{\alpha\beta}^{\sigma_{1}\sigma_{2}}(\bm{k})
\sum_{m_{1}m_{2}m_{3}m_{4}}
\cr&\quad\times
(X^{i\dagger})_{m_{1}m_{2}}^{\sigma_{2}\sigma_{1}}
[U_{\sigma_{1}}(\bm{k})]_{m_{2}\alpha}[U_{\sigma_{1}}^{*}(\bm{k})]_{m_{3}\alpha}
\cr&\quad\times
(X^{j})_{m_{3}m_{4}}^{\sigma_{1}\sigma_{2}}
[U_{\sigma_{2}}^{*}(\bm{-k})]_{m_{4}\beta}[U_{\sigma_{2}}(\bm{-k})]_{m_{1}\beta}
,
\end{align}
with
\begin{align}
I_{\alpha\beta}^{\sigma_{1}\sigma_{2}}(\bm{k})&=T\sum_{n}G_{\alpha\sigma_{1}}^{(0)}(\bm{k},i\omega_{n})G_{\beta\sigma_{2}}^{(0)}(-\bm{k},-i\omega_{n})
\cr&
=\frac{\tanh(\beta\xi_{\bm{k}\alpha\sigma_{1}}/2)+\tanh(\beta\xi_{-\bm{k}\beta\sigma_{2}}/2)}{2(\xi_{\bm{k}\alpha\sigma_{1}}+\xi_{-\bm{k}\beta\sigma_{2}})},
\end{align}
where $G_{\alpha\sigma}^{(0)}(\bm{k},i\omega_{n})=1/(i\omega_n-\xi_{\bm{k}\alpha\sigma})$.

The linearized gap equations are explicitly given by
\begin{align}
\begin{pmatrix} \psi \\ D_{z} \\ d_{z} \end{pmatrix}
&=-
\begin{pmatrix}
M_{11} & M_{12} & M_{13} \\
M_{21} & M_{22} & M_{23} \\
M_{31} & M_{32} & M_{33}
\end{pmatrix}
\begin{pmatrix} \psi \\ D_{z} \\ d_{z} \end{pmatrix},
\label{a1gap}
\\&
M_{11}=\frac{1}{2}(UK_{11}^{(0)}+J_{0}K_{21}^{(0)}),
\\&
M_{22}=\frac{1}{4}(U+J_{2})K_{22}^{(0)}+\frac{1}{2}J_{0}K_{21}^{(0)},
\\&
M_{33}=\frac{1}{4}(U-3J_{2})K_{33}^{(0)},
\\&
M_{12}=\frac{1}{2}(J_{0}K_{22}^{(0)}+UK_{21}^{(0)}),
\\&
M_{21}=\frac{1}{2}J_{0}K_{11}^{(0)}+\frac{1}{4}(U+J_{2})K_{21}^{(0)},
\\&
M_{13}=\frac{1}{2}(UK_{13}^{(0)}+J_{0}K_{23}^{(0)}),
\\&
M_{31}=\frac{1}{4}(U-3J_{2})K_{23}^{(0)},
\\&
M_{23}=\frac{1}{2}J_{0}K_{13}^{(0)}+\frac{1}{4}(U+J_{2})K_{23}^{(0)},
\\&
M_{32}=\frac{1}{4}(U-3J_{2})K_{13}^{(0)},
\label{m32}
\\
d_{x,y}&=-\frac{1}{4}(U-3J_{2})K_{44}^{(0)}d_{x,y},
\\
D_{x,y}&=-\frac{1}{4}(U-J_{2})K_{66}^{(0)}D_{x,y}.
\label{egap}
\end{align}

These results are easily understood by the group theoretical argument.
In the presence of the Ising-type SOC, the symmetry operations must apply to both the orbital and spin spaces simultaneously, and the components are classified by the irreducible representations for the spin-orbital state.
As shown in Table~\ref{tab:1}, $\psi$, $D_{z}$, and $d_{z}$ belong to the same irreducible representation $\mathrm{A}_{1}'$ in the $D_{3h}$ point group.
Therefore, they mix with each other as Eq.~(\ref{a1gap}).
The other components are discriminated by the different irreducible representations, E$'$ and E$''$.

When we turn off the Ising-type SOC, the symmetry operations are defined separately in the orbital and spin spaces.
In other words, the components are classified by the orbital irreducible representation (o-irrep.) and the spin magnitude $S$.
As shown in Table~\ref{tab:1}, $(\psi,D_{z})$ and $d_{z}$ belong to the different o-irreps., A$'_{1}$ and A$'_{2}$, and hence they do not mix with each other.
Moreover, $d_{z}$ and $(d_{x},d_{y})$ must be degenerate, since they constitute the components of the spin-triplet $S=1$.
These statements are explicitly confirmed by the facts that $K_{13}^{(0)}=K_{23}^{(0)}=0$ and $K_{33}^{(0)}=K_{44}^{(0)}$ for $\lambda=0$.

These gap equations will be solved with the fixed electron density, $n$, which is obtained from
\begin{align}
n=\frac{T}{N_{0}}\sum_{\bm{k}n}\sum_{m\sigma}[G(\bm{k},i\omega_{n})]^{\sigma\sigma}_{mm}e^{i\omega_{n}0_{+}}.
\end{align}

\section{Results and discussions}

We elucidate the transition temperature $T_{\rm c}$ and the ratio of the pairing components at $T_{\rm c}$ by solving the linearized gap equations, Eqs.~(\ref{a1gap})-(\ref{egap}).
$T_{\rm c}$ is determined by the condition that the maximum eigenvalue of the kernel in the linearized gap equation reaches unity, and its eigenvector provides the ratio of the pairing components at $T_{\rm c}$.
We fix the leading attractive interaction $U=-0.5$ eV, and we use the $\bm{k}$-mesh as $N_{0}=900\times 900$.

\begin{figure}[t!]
\begin{center}
\includegraphics[width=8.5 cm]{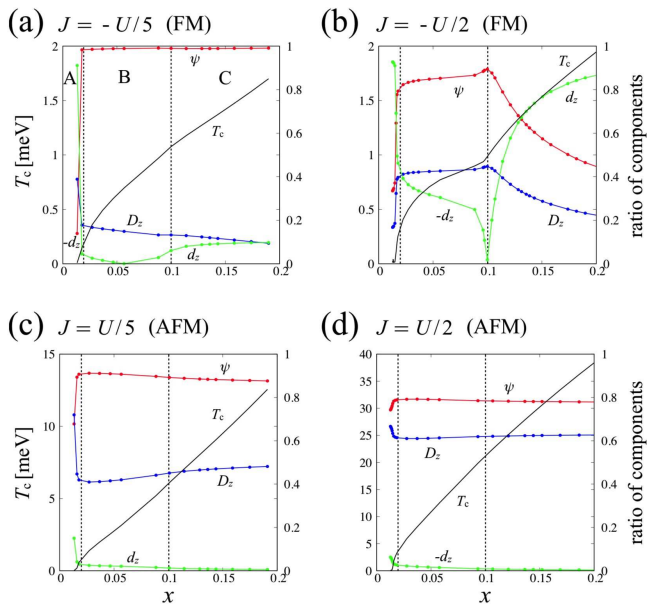}
\caption{
Doping dependences of $T_{\rm c}$ and the Cooper-pair components for $U=-0.5$ eV.
(a) $J=-U/5$ (ferromagnetic), (b) $J=-U/2$, (c) $J=U/5$ (antiferromagnetic), and (d) $J=U/2$.
The black line (left axis) represents $T_{\rm c}$, and the red, blue, and green lines (right axis) represent the components, $\psi$ (SS), $D_{z}$ (SS-OT), and $d_{z}$ (ST-OS), respectively.
The pairing states $(d_{x},d_{y})$ and $(D_{x},D_{y})$ have much lower $T_{\rm c}$ (not shown).
The vertical dotted lines indicate the doping level at which the Fermi-surface topology changes.
Note that $d_{z}$ changes its sign at the boundary between A and B for (a), and between B and C for (b).
}
\label{fig:3}
\end{center}
\end{figure}

Figure~\ref{fig:3} shows the doping dependences of $T_{\rm c}$ and the ratio of the Cooper-pair components for (a) $J=-U/5$ (ferromagnetic), (b) $J=-U/2$, (c) $J=U/5$ (antiferromagentic), and (d) $J=U/2$.
In most cases, the conventional spin-singlet pairing $\psi$ in the $\Gamma$ pocket dominates the pairing, since the DOS of the $\ket{0}$ orbital near $\Gamma$ point are larger than those of the $\ket{\pm2}$ orbitals near K and K$'$ points.
The $(d_{x},d_{y})$ and $(D_{x},D_{y})$ pairings do not appear as $T_{\rm c}$ is much lower than that of the ${\rm A}'_{1}$ state.
At low dopings in the region A, $T_{\rm c}$ is exponentially low due to small DOS of the upper spin-split $\ket{\pm2}$ bands.
Indeed, we estimate $T_{\rm c}$ according to the BCS formula for the critical temperature, and $T_{\rm c}$ in the region A is one or two orders of magnitude lower than that in the region B.
Note that the exchange interaction $J_{0}$ enhances $T_{\rm c}$ of the ${\rm A}'_{1}$ state by the pair scattering between the $\ket{0}$ and $\ket{\pm2}$ pockets, which is common mechanism of enhancing $T_{\rm c}$ in multi-gap superconductivity~\cite{Suhl, Kondo, Soda}.
However, the antiferromagnetic exchange interaction itself suppresses the attractive interaction for $d_{z}$ and enhances for $D_{z}$ as was mentioned (see also Eqs.~(\ref{v12}) and (\ref{v345})).
Therefore, the detailed balance of the parameters gives the highest $T_{\rm c}$ for the ${\rm A}'_{1}$ state.
The relative sign between $(\psi, D_{z})$ and $d_{z}$ depends on the sign of $U-3J$ as implied in $M_{32}$ of Eq.~(\ref{m32}).

As doping increases, $T_{\rm c}$ monotonously increases.
Especially, in the case of Fig.~\ref{fig:3}(b), $T_{\rm c}$ is remarkably enhanced in the region C, where the Fermi surfaces of the lower split bands around K and K$'$ points appear.
This is because the strongest attraction of the $d_{z}$ pairing works efficiently when both the spin-split bands are involved.

 \begin{figure}[t!]
\begin{center}
\includegraphics[width=8.5cm]{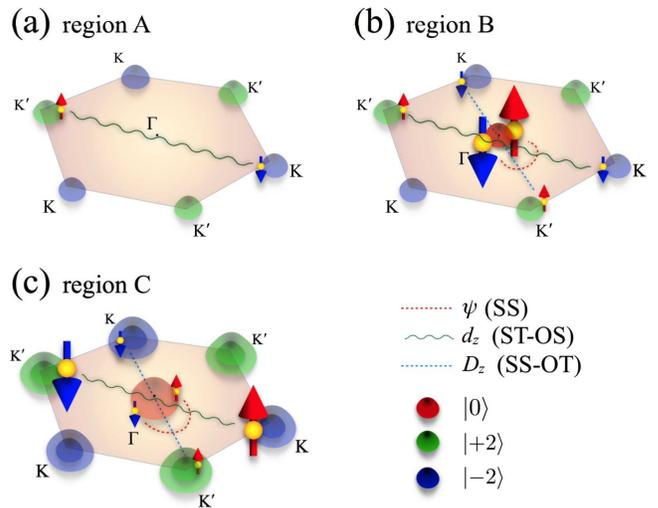}
\caption{
Schematic illustrations for the representative pairing states for $U=-0.5$ eV and $J=-U/2$ corresponding to the case in Fig.~\ref{fig:3}(b), (a) in the region A, (b) in the region B, and (c) in the region C.
The colored areas with red, green, blue indicate the spin-split hole pockets of the $\ket{0}$, $\ket{+2}$, and $\ket{-2}$ orbitals, respectively.
The size of the arrows roughly indicates the magnitude of the pairing components.
}
\label{fig:4}
\end{center}
\end{figure}

Schematic illustrations for the representative pairing states in the case of Fig.~\ref{fig:3}(b) are shown in Fig.~\ref{fig:4}.
In the region A, the $d_{z}$ (ST-OS) pairing within the upper split bands around K and K$'$ points occurs with very small $T_{\rm c}$.
In the region B, the $\psi$ (SS) pairing in almost non-split bands having relatively large DOS around $\Gamma$ point dominates over the $d_{z}$ and $D_{z}$ pairings.
The small but finite weights of $d_{z}$ and $D_{z}$ are favorable to gain the Cooper-pair hopping energy as similar to the multigap superconductivity.
In the region C, the weights of $\psi$ and $d_{z}$ become comparable and the latter dominates over the former as $x$ further increases.
This is because both the upper and lower split bands near K and K$'$ points contribute to the $d_{z}$ pairing, which has the strongest attraction.

So far, no superconducting states have been observed in the hole-doped monolayer MoS$_{2}$ in contrast to the case of the electron-doping.
The smallness of the DOS of the upper split bands around K and K$'$ points at small doping rates may be one of the reasons why the hole-doped superconductivity does not emerge.
If it were a main reason, the further doping would bring about the superconducting state of ${\rm A}'_{1}$ symmetry.
The resulting pairing states have a mixture of the spin singlet and triplet, in which the predominant components depend on the sign of the exchange interactions.

\section{Summary}

We have investigated the Cooper-pair symmetry of the hole-doped monolayer MoS$_{2}$.
The electronic structure of the valence band edge of monolayer MoS$_{2}$ is characterized predominantly by Mo $d_{z^{2}}$, $d_{x^{2}-y^{2}}$, and $d_{xy}$ orbitals.
The hole pocket near $\Gamma$ point is characterized by almost spin-degenerate $d_{z^{2}}$ orbital, while the pockets near K and K$'$ points consist of the spin-split $d_{x^{2}-y^{2}}$ and $d_{xy}$ orbitals due to the Ising-type SOC.
As the doping rate increases, the hole pockets appear first in the upper split bands near K and K$'$ points, and subsequently appear in the degenerate band near $\Gamma$ point, and the lower split bands near K and K$'$ points.
The DOS of the degenerate band is larger than those of the split bands.

Reflecting the above electronic structure, we have found the dominant SS pairing ($\psi$) in whole doping range having the $\Gamma$ pocket.
For the lower doping where the $\Gamma$ pocket disappears, $T_{\rm c}$ is found to be too low due to the small DOS of the upper split bands.
For higher doping rate where several Fermi pockets appear, the SS pairing has a mixture of the SS-OT ($D_{z}$) and ST-OS ($d_{z}$) pairings belonging to A$_{1}'$ symmetry.
The mixing is caused by the exchange interactions, which usually enhance $T_{\rm c}$ by 
the inter-band proximity effect.
Moreover, the ferromagnetic exchange interactions considerably increase the weight of the ST-OS pairing.
It even dominates over the SS pairing for moderately large exchange interactions at high doping rate.

In spite of these fascinating characteristics, so far no superconducting states have been realized in hole-doped monolayer MoS$_2$.
However, a series of compounds such as TaS$_{2}$ and NbSe$_{2}$, which has a band structure similar to that of MoS$_{2}$, exhibits superconductivity, and their in-plane $H_{\rm c2}$ are much higher than the Pauli limit~\cite{Xi,Barrera}.
Thus, it is interesting to discuss the connection between these observations and the present work, which is left for future investigation.

\begin{acknowledgments}
The authors would like to thank T. Arima and T. Nojima for fruitful discussions.
This research was supported by JSPS KAKENHI Grants Numbers 15K05176 and 18H04296 (J-Physics). 
\end{acknowledgments}


\end{document}